# Evaluating residual acceleration noise for the TianQin gravitational waves observatory with an empirical magnetic field model

Wei Su,[1,2,*] Ze-Bing Zhou,[2,†] Yan Wang,[2] Chen Zhou,[3] P. F. Chen,[4,5,‡] Wei Hong,[2] J. H. Peng,[1] Yun Yang,[6] and Y. W. Ni[4,5]

[1]*TianQin Research Center for Gravitational Physics & School of Physics and Astronomy, Sun Yat-sen University (Zhuhai Campus), Zhuhai 519082, People's Republic of China*
[2]*MOE Key Laboratory of Fundamental Physical Quantities Measurements, Hubei Key Laboratory of Gravitation and Quantum Physics, PGMF, Department of Astronomy, and School of Physics, Huazhong University of Science and Technology, Wuhan 430074, China*
[3]*Department of Space Physics, School of Electronic Information, Wuhan University, Wuhan 430072, China*
[4]*School of Astronomy and Space Science, Nanjing University, Nanjing 210023, China*
[5]*Key Laboratory of Modern Astronomy & Astrophysics, Nanjing University, Nanjing 210023, China*
[6]*School of Mathematical Sciences, Nanjing Normal University, Nanjing, 210023, People's Republic of China*

TianQin (TQ) project plans to deploy three satellites in space around the Earth to measure the displacement change of test masses caused by gravitational waves via laser interferometry. The requirement of the acceleration noise of the test mass is on the order of $10^{-15}$ m s$^{-2}$ Hz$^{-1/2}$ in the sensitive frequency range of TQ, which is so stringent that the acceleration noise caused by the interaction of the space magnetic field with the test mass needs to be investigated. In this work, by using the Tsyganenko model, a data-based empirical space magnetic field model, we obtain the magnetic field distribution around TQ's orbit spanning two solar cycles in 23 years from 1998 to 2020. With the obtained space magnetic field, we derive the distribution and amplitude spectral densities of the acceleration noise of TQ in 23 years. Our results reveal that the average values of the ratio of the acceleration noise caused by the space magnetic field to the requirements of TQ at 1 mHz ($R_{1\,\mathrm{mHz}}$) and 6 mHz ($R_{6\,\mathrm{mHz}}$) are $0.123 \pm 0.052$ and $0.027 \pm 0.013$, respectively. The occurrence probabilities of $R_{1\,\mathrm{mHz}} > 0.2$ and $> 0.3$ are only 7.9% and 1.2%, respectively, and $R_{6\,\mathrm{mHz}}$ never exceeds 0.2.

## I. INTRODUCTION

To date, nearly 100 gravitational wave (GW) events have been detected, all by ground-based GW detectors, e.g., advanced LIGO, advance Virgo [1,2], and KAGRA [3]. The sensitive frequency band of the ground-based GW detectors are in the range of 10–1000 Hz. In order to expand the sensitive frequency band of GW detection to lower bands (0.1 mHz–1 Hz), several space-borne projects have been proposed, e.g., LISA [4], gLISA [5], Taiji [ALIA descoped; 6], TianQin [TQ; 7], ASTROD-GW [8], BBO [9], and DECIGO [10]. While some of them are in the heliocentric orbits, others are in the geocentric orbits.

TQ is a geocentric space-borne GW detector with an orbital altitude of $10^5$ km measured from the geocenter [11–13]. It is composed of three satellites, which are connected by laser links with each arm length of about $1.7 \times 10^5$ km [14–16]. The technical principle of the space-borne GW missions is to use laser interferometry to measure the displacement change between two test masses (TMs) due to GWs. For space-borne GW detectors, the requirement of measurement accuracy is extremely stringent. Taking TQ as an example, the relative displacement and acceleration accuracies are required to be on the order of $10^{-12}$ m Hz$^{-1/2}$ and $10^{-15}$ m s$^{-2}$ Hz$^{-1/2}$ in mHz, respectively [7,17].

It is noted that neither the heliocentric GW detectors (e.g., LISA and Taiji) nor the geocentric ones (e.g., TQ) are embedded in vacuum. Instead, the space between the TMs is filled with solar wind plasma, magnetic field, cosmic rays, and so on. Laser propagation in space plasma can cause optical path difference noise for both LISA [18] and TQ [19,20], and solar wind can disturb the spacecraft of GW detection [21]. In addition, remanent magnetic moment, residual charge, and the finite magnetic susceptibility would all cause the TMs to interact with space magnetic field, causing magnetic torque and Lorentz force [22,23], resulting in acceleration noise [24,25]. Subsequently, the acceleration noise caused by interplanetary magnetic field was investigated for LISA, LISA PathFinder, and eLISA [26–28]. Recently, spacecraft and interplanetary magnetic environment of LISA Pathfinder has been reported based on the in-flight measurements [29]. For TQ, the residual acceleration caused by the space magnetic field was studies [30], and there have been continuous advances in the technologies and fabricating of the TMs, e.g., mass distribution, magnetic susceptibility, and charge management [31–34].

[*]suwei25@mail.sysu.edu.cn
[†]zhouzb@hust.edu.cn
[‡]chenpf@nju.edu.cn

The solar surface is permeated with small-scale magnetic elements and peppered with sunspots. Whereas some magnetic field lines close in the corona, forming coronal loops, other magnetic field lines extend all the way out to the interplanetary space, forming the environment of the space weather around the Earth [35]. The solar activities resulting from the changing magnetic field in the solar atmosphere are the dominant factor leading to the magnetic field and plasma variations around the orbits of the space-borne GW detectors. In a previous work [30], we used a global magnetohydrodynamic (MHD) simulation, Space Weather Modeling Framework (SWMF, [36]), with the input of the real time solar wind data to get the temporal and spatial distributions of space plasma and magnetic field around the orbit of TQ. Based on the magnetic field obtained from the MHD simulation, we calculated the acceleration noise caused by the space magnetic field, and derived the amplitude spectral densities (ASDs) of the acceleration noise in one TQ orbit [30]. However, in that paper, we only studied a single case with typical input parameters (e.g., solar wind dynamic pressure, and interplanetary magnetic field), and due to computational resources limitations, we did not conduct a statistical study with the SWMF model. In this work, we expect to obtain statistical results for acceleration noise due to the space magnetic field in more than a solar cycle (11 years). During a solar cycle, there are heliophysical phenomena on different timescales [37], e.g., MHD instabilities [38–40] and solar eruptions [41–43] on the order of days to hours, and plasma waves [44–48] on the order of seconds to subseconds. The probabilities of these phenomena in the solar maximum year and minimum year are significantly different, for example, the solar wind speed near the Earth is inversely correlated to the sunspot number, while the solar eruptions are positively correlated to the sunspot number. However, it is noticed that MHD numerical simulations are too time consuming, for example, it would take more than a hundred years to obtain the space magnetic field during one solar cycle by using the MHD method as used in Su *et al.* [30]. In order to evaluate the level of acceleration noise under different solar wind conditions during more than one solar cycle, we herein use a semiempirical model of the Earth's magnetic field, the Tsyganenko model [49,50], to calculate the acceleration noise of the proposed TQ mission due to the space magnetic field.

This paper is organized as follows: Sec. II introduces the analytical model of acceleration caused by the magnetic field. Section III presents the model of the Earth's magnetic field, i.e., the Tsyganenko model. The results are presented in Sec. IV, which are discussed in Sec. V, with a summary in Sec. VI.

## II. ACCELERATION NOISE DUE TO SPACE MAGNETIC FIELD FOR GW DETECTIONS

Despite all the demagnetization processing, the magnetic moment of a TM is not zero in reality. As a result, the magnetostatic force on a TM with magnetic moment ($M_{\rm tm}$) in magnetic field ($B$) can be expressed as

$$F = \nabla(M_{\rm tm} \cdot B). \quad (1)$$

Here, $B$ can be separated to the spacecraft magnetic field $B_{\rm sc}$ and the space magnetic field $B_{\rm sp}$. $M_{\rm tm}$ can be separated to the inductive magnetic moment $M_{\rm i}$ and the remanent magnetic moment $M_{\rm r}$. In addition, $M_{\rm i}$ is composed of two components which are induced by $B_{\rm sc}$ and by $B_{\rm sp}$, denoted as $M_{\rm isc}$ and $M_{\rm isp}$, respectively. With all these notations, the acceleration of the TM due to the magnetic field can be written as

$$a_{\rm M} = \frac{1}{m}\nabla[(M_{\rm r} + M_{\rm isp} + M_{\rm isc}) \cdot (B_{\rm sp} + B_{\rm sc})], \quad (2)$$

here, $m$ is the mass of the TM; $M_{\rm isc} = \chi_{\rm m} V_{\rm tm} B_{\rm sc}/\mu_0$ and $M_{\rm isp} = \chi_{\rm m} V_{\rm tm} B_{\rm sp}/\mu_0$, where $\chi_{\rm m}$ is the magnetic susceptibility, $V_{\rm tm}$ is the volume of the TM, and $\mu_0$ is the vacuum magnetic permeability.

Because the TM is enclosed by the inertial sensor housing in the disturbance reduction system, the electric current inside the TM is neglected here, so the Ampère-Maxwell current law contains the displacement current term only here. Considering the magnetic and electric shielding of the TM housing, we introduce the magnetic and electric shielding factors, which are denoted as $\xi_m$ and $\xi_e$. Combining the vector operation rules, Eq. (2) can be expanded as follows (see Sec. 2 of [30] for more details):

$$\begin{cases} a_{\rm M1} = \frac{1}{m\xi_m}[(M_{\rm r} + 2M_{\rm isp}) \cdot \nabla]B_{\rm sc} \\ a_{\rm M2} = \frac{1}{m\xi_m}[(M_{\rm r} + 2M_{\rm isc}) \cdot \nabla]B_{\rm sp} \\ a_{\rm M3} = \frac{1}{m\xi_e}(M_{\rm r} + 2M_{\rm isc}) \times \frac{\varepsilon_0\mu_0\partial E_{\rm sp}}{\partial t} \\ a_{\rm M4} = \frac{2}{m\xi_m}M_{\rm isp}\nabla B_{\rm sp} \\ a_{\rm M5} = \frac{1}{m\xi_e}(M_{\rm r} + 2M_{\rm isp}) \times \frac{\varepsilon_0\mu_0\partial E_{\rm sc}}{\partial t} \\ a_{\rm M6} = \frac{2}{m\xi_m}M_{\rm isc}\nabla B_{\rm sc} \end{cases}, \quad (3)$$

where $\varepsilon_0$ is the vacuum electric permittivity, $\boldsymbol{E}_{\rm sp}$ is the space electric field, and $\boldsymbol{E}_{\rm sc}$ is the electric field of spacecraft of the TM.

In this paper, the parameters of the TM are taken as follows: $m = 2.45$ kg, $|\boldsymbol{M}_{\rm r}| = 20$ nA m$^2$, the magnetic susceptibility of the Pt-Au alloy is $\chi_m = 10^{-5}$, the size of the TM cube is $5 \times 5 \times 5$ cm$^3$ [7], both $\xi_m$ and $\xi_e$ are set to be 10. The values of all these parameters are the same as in Su *et al.* [30].

Considering the typical magnitudes of $\boldsymbol{B}_{\rm sp}$, $\nabla \boldsymbol{B}_{\rm sp}$, and the temporal fluctuations of $\boldsymbol{E}_{\rm sp}$ around the TQ's orbit as 10 nT, 0.01 nT m$^{-1}$, and $10^{-4}$ V m$^{-1}$ s$^{-1}$, the magnitudes of $\boldsymbol{a}_{\rm M1}$, $\boldsymbol{a}_{\rm M2}$, $\boldsymbol{a}_{\rm M3}$, and $\boldsymbol{a}_{\rm M4}$ are on the order of $10^{-14}$, $10^{-20}$, $10^{-29}$, and $10^{-22}$ m s$^{-2}$, respectively [30]. Taking the typical magnitude of temporal variations of $\nabla \boldsymbol{B}_{\rm sc}$ and $\boldsymbol{E}_{\rm sc}$ (denoted as $\delta \nabla \boldsymbol{B}_{\rm sc}$ and $\delta \boldsymbol{E}_{\rm sc}$) as 1 nT m$^{-1}$ Hz$^{-1/2}$ and 100 μV m$^{-1}$ Hz$^{-1/2}$ at 1 mHz [25,51], the magnitude of $\boldsymbol{a}_{\rm M5}$ and $\boldsymbol{a}_{\rm M6}$ are on the order of $10^{-32}$ and $10^{-18}$ m s$^{-2}$ at 1 mHz, and the fluctuation of $\boldsymbol{a}_{\rm M1}$ (denoted as $\boldsymbol{a}_{\rm M1}^{\delta \nabla \boldsymbol{B}_{\rm sc}}$) due to the temporal variations of $\nabla \boldsymbol{B}_{\rm sc}$ is on the order of $10^{-19}$ m s$^{-2}$, which are much smaller than $\boldsymbol{a}_{\rm M1}$. Thus, in this work, we neglect $\boldsymbol{a}_{\rm M2}$, $\boldsymbol{a}_{\rm M3}$, $\boldsymbol{a}_{\rm M4}$, $\boldsymbol{a}_{\rm M5}$, and $\boldsymbol{a}_{\rm M6}$. And since $\boldsymbol{a}_{\rm M1}^{\delta \nabla \boldsymbol{B}_{\rm sc}} \ll \boldsymbol{a}_{\rm M1}$, we also neglect $\boldsymbol{a}_{\rm M1}^{\delta \nabla \boldsymbol{B}_{\rm sc}}$ and take $\nabla \boldsymbol{B}_{\rm sc}$ as a constant. Consider that the variation of $\boldsymbol{M}_{\rm r}$ is much slower than that of the inductive magnetic moment caused by $\boldsymbol{B}_{\rm sp}$ ($\boldsymbol{M}_{\rm isp}$), $\boldsymbol{M}_{\rm r}$ is taken as a constant here. Thus, in $\boldsymbol{a}_{\rm M1}$, the term $(2/m\xi_m)(\boldsymbol{M}_{\rm r} \cdot \nabla)\boldsymbol{B}_{\rm sc}$ is a constant, that is to say, $(2/m\xi_m)(\boldsymbol{M}_{\rm r} \cdot \nabla)\boldsymbol{B}_{\rm sc}$ in $\boldsymbol{a}_{\rm M1}$ is a direct current (dc) term, which has no effect on the spectral density. Hereafter, we remove the dc term in $\boldsymbol{a}_{\rm M1}$ and focus on the acceleration noise caused by the space magnetic field ($\boldsymbol{B}_{\rm sp}$), and $\boldsymbol{a}_{\rm M1}$ can be rewritten as

$$\boldsymbol{a}_{\rm M1} = \frac{2}{m\xi_m}(\boldsymbol{M}_{\rm isp} \cdot \nabla)\boldsymbol{B}_{\rm sc}. \tag{4}$$

The magnitude of $\boldsymbol{a}_{\rm M1}$ is on the order of $10^{-16}$ m s$^{-2}$, which is consistent with the fluctuation of $\boldsymbol{a}_{\rm M1}$ in Su *et al.* [30].

The TM housing might be penetrated by energetic particles, e.g., solar energetic particles (SEPs) and galactic cosmic rays (GCRs), therefore the TMs might be charged [52]. There would be Lorentz force on the charged TMs in the background space magnetic field. Meanwhile, the Lorentz force on the TM can be partially compensated by the Hall voltage when the metallic enclosure of the TM travels through the space magnetic field, which effectively acts as a shield [23]. The effective shielding coefficient $\eta$ was taken to be 0.03 in Sumner *et al.* [23], while in this work, $\eta$ is taken to be 0.1 for a conservative estimate. Thus, the acceleration due to the Lorentz force on the TM with residual charge ($q$) can be expressed as

$$\boldsymbol{a}_{\rm L} = \frac{\eta}{m} q \boldsymbol{v} \times \boldsymbol{B}_{\rm sp}, \tag{5}$$

where $\boldsymbol{v}$ is the speed of the TM.

### III. TSYGANENKO MODEL

MHD numerical simulations are an important method in the study of heliophysics and have been widely applied to investigate various phenomena in the heliosphere [53–55]. The SWMF model was developed by Tóth *et al.* [36], which can simulate the interaction between the solar wind and the Earth's magnetosphere by solving MHD equations with the real-time observations of the solar wind as input conditions and the Earth's magnetic field as the boundary conditions. The code can obtain the evolution of the Earth's magnetosphere in response to the solar wind. Su *et al.* [30] applied the interplanetary magnetic field obtained from the SWMF simulations to study the acceleration noise due to the space magnetic field. However, the MHD simulations require too many computational resources. If we had used MHD simulations to investigate the evolution of the Earth's magnetic field in one solar cycle (11 years) with a time resolution of 1 min, it would spend hundreds of years on simulations even with supercomputers. Therefore, it is not practical to use MHD simulations in investigating the secular evolution of the Earth's magnetosphere.

On the other hand, the Tsyganenko magnetic field model is a data-based empirical model, which can calculate the Earth's magnetic field much faster than MHD simulations [56,57]. The inner boundary of the Tsyganenko model is provided by the international geomagnetic reference field model, and the input parameters for the outer boundary conditions are the solar wind dynamic pressure ($P_{\rm dyn}$), the magnetic inclination of the Earth, $B_z$ and $B_y$ components of the interplanetary magnetic field, the $D$st index, etc. [49,50,58,59]. The output of the Tsyganenko model are $B_x$, $B_y$, and $B_z$ components of the Earth's magnetic field in the geocentric solar magnetospheric coordinates. $P_{\rm dyn}$ is the primary factor in shaping the structure of the Earth's magnetosphere. The variation of the angle between the magnetic inclination of the Earth and the Sun-Earth line can affect the morphology of the Earth's magnetosphere [60]. The interplanetary magnetic field $B_z$ is a key factor that affects the exchange of energy and plasma between the Earth's magnetosphere and the solar wind [61]. The interplanetary magnetic field $B_y$ causes the dawn-dusk asymmetry of the magnetosphere [62]. $D$st index characterizes the intensity of the ring current of the magnetosphere, where the ring current is an important ingredient in the Earth's current system [63,64].

One example of the magnetic field around the Earth calculated by the Tsyganenko model is shown in Fig. 1. On the dayside, the solar wind compresses the magnetosphere and forms the magnetopause; on the nightside, the solar wind stretches the Earth's magnetic field and forms an antiparallel magnetic field structure with a current sheet in the magnetotail, which is consistent with the *in situ* observations over decades [65–67]. The Tsyganenko model has been validated in various solar wind conditions, even in the presence of magnetic storms [50]. So far, the Tsyganenko model has been widely applied in space physics and space science research [68].

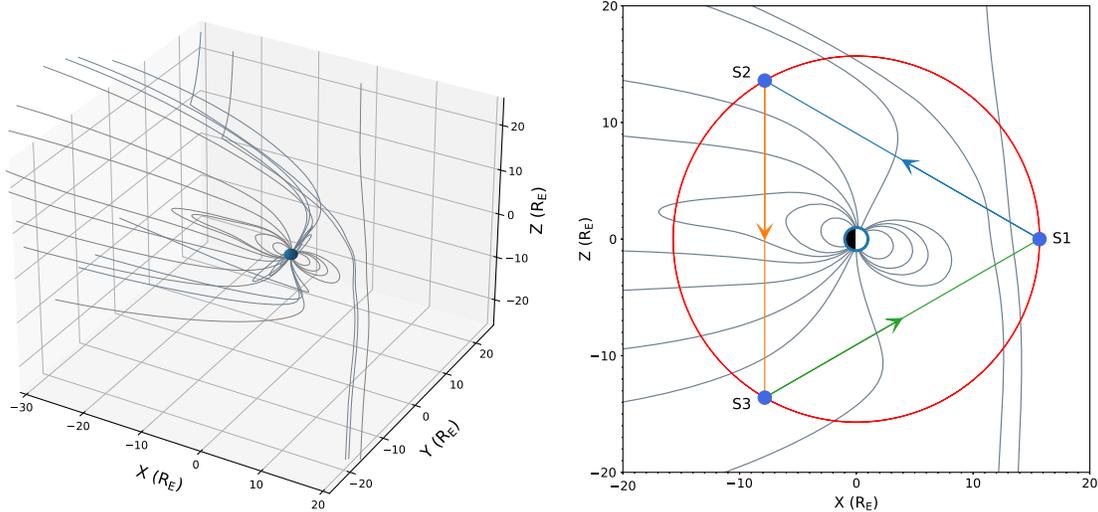

FIG. 1. Schematic of TQ satellites in the background of the Earth magnetic field. The magnetic field lines calculated by the Tsyganenko model are shown as grey lines in both panels. Left panel displays the 3D magnetic field around the Earth. In right panel, the background is the magnetic field in a 2D plane, the red circle is the orbit of TQ, the three blue dots are the TQ satellites, and the laser links S1–S2, S2–S3, and S3–S1 are represented by blue, orange, and green lines, respectively.

## IV. RESULTS

### A. Magnetic field

In this paper, we take the OMNI solar wind data as the input of the Tsyganenko model. OMNI is a database of the magnetic field, plasma, and energetic particles relevant to heliospheric studies, which includes *in situ* measurements data from multiple satellites [69]. The OMNI data contain the input parameters for the Tsyganenko model (magnetic field, $P_{\rm dyn}$, and $D$st index) and other parameters (such as number density and temperature of electrons and protons, sunspot number, etc.). For this work, we take the OMNI data in a time range of 23 years from the beginning of 1998 to the end of 2020. Since a solar cycle is about 11 years, the time interval of the data we are analyzing here covers two solar cycles.

The time resolution of the input data we use is 1 minute for $P_{\rm dyn}$ and magnetic field data, while for the $D$st index the time resolution is 1 hour. We use the spline interpolation to generate $D$st data with 1 minute time cadence. Taking six cycles of a TQ satellite around the Earth as examples, the distributions of the inputs of the Tsyganenko model are shown in Fig. 2. Here, the initial orbit position of a TQ satellite (S1) at 00∶00∶00 UT on January 1 of each year is set as (0, $10^5$, 0) km in the geocentric solar ecliptic coordinates. Considering the period of TQ satellites around the Earth is about 3.65 days [12,13], we can get 100 complete orbits of the TQ satellites around the Earth each year, meaning 2300 orbits in 23 years. Combining the TQ orbit and the input parameters, we use the Tsyganenko model to get the space magnetic field evolution along the 2300 orbits of a TQ satellite around the Earth. Here, we take six orbits of the TQ satellite as examples, the evolution of the space magnetic field as the satellite move in six orbits are shown in the six panels of Fig. 3, respectively.

### B. Acceleration

As a vector, acceleration has three components. In Su *et al.* [30], we calculated the absolute value of the acceleration noise due to the space magnetic field without distinguishing between the sensitive axis and the nonsensitive axes. Since the absolute value of the acceleration noise is greater than or equal to the component of acceleration noise on the sensitive axis, the acceleration noise was overestimated in Su *et al.* [30]. The acceleration noise requirement for TQ on the sensitive axis is on the order of $10^{-15}$ m s$^{-2}$ Hz$^{-1/2}$, which is about two orders of magnitude lower than that on the nonsensitive axes, so the acceleration noise on the sensitive axis deserves special attention.

In this work, we distinguish between sensitive axis and insensitive axes for each laser link. Note that the sensitive axis is meaningful for one laser link, which is along the link direction. The right panel of Fig. 1 illustrates the schematic of the TQ satellites in the background space magnetic field. The three satellites of TQ (denote as S1, S2, and S3 in the figure) are forming an equilateral triangle and moving around the Earth, they are connected to each other by three laser links. What we need to measure by laser interferometry is the variation of the displacement between the TMs on two satellites, that is, the sensitive axis is along the laser link. Here, we take a TM in S1 as an example to study the acceleration noise due to space magnetic field. In addition, for the link S1–S2, we assign the S1 → S2 direction to be positive which is shown as a blue line with an arrow in the right panel of Fig. 1.

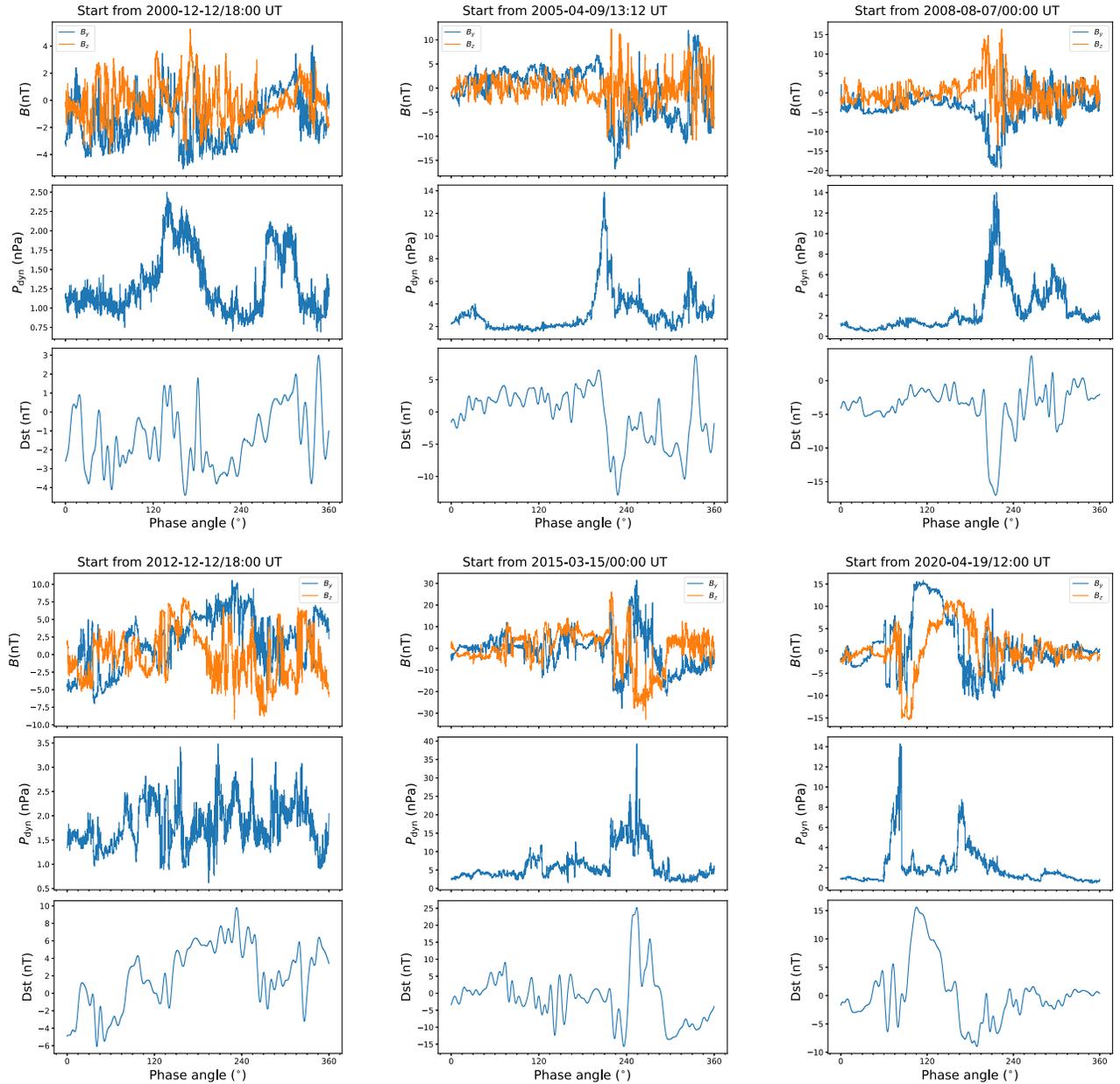

FIG. 2. Input parameters ($B_y$, $B_z$, solar wind dynamic pressure $P_{\rm dyn}$, and $D$st index) of the Tsyganenko model, the six panels are the distributions of the inputs along each orbit of the TQ satellite in the six cases.

We use the Tsyganenko model to obtain the components of the space magnetic field ($B_x$, $B_y$, $B_z$) along the orbit of TQ, and substitute ($B_x$, $B_y$, $B_z$) into Eqs. (4) and (5), $a_{\rm M1}$ and $a_{\rm L}$ are calculated. What we need, i.e., the components of $a_{\rm M1}$ and $a_{\rm L}$ along the sensitive axis, can be obtained. Here, the components of $a_{\rm M1}$ and $a_{\rm L}$ along the sensitive axis are marked as $A_M$ and $A_L$, respectively.

We get the distributions of $A_M$ and $A_L$ along 2300 orbital cycles of S1 around the Earth. In Fig. 4, we display $A_M$ and $A_L$ distributions in the six cases, where the blue lines correspond to $A_M$, and the orange lines correspond to $A_L$. The magnitudes of $A_M$ and $A_L$ are on the order of $0.1 \times 10^{-15}$ and $0.01 \times 10^{-15}$ m s$^{-2}$, respectively. It is seen that $A_M$ is about one order of magnitude larger than $A_L$, therefore $A_M$ is the dominate source of acceleration noise due to the space magnetic field.

Furthermore, we calculate the ASD of $A_M$ for the 2300 cycles of S1 around the Earth, and then apply the Savitzky-Golay filter [70] to smooth the ASD profiles. The smoothed ASD profiles of $A_M$ in the six cases are shown as the blue curves in Fig. 5. We fit the ASD profiles of $A_M$ with the power-law functions $A_{1\,\mathrm{mHz}}(f/1\,\mathrm{mHz})^\alpha$, here $A_{1\,\mathrm{mHz}}$ is the fitted value of the ASD of $A_M$ at 1 mHz, $f$ is the frequency, $\alpha$ is the spectral index of the fitted profile. The fitted spectra are represented as the red dashed lines in Fig. 5.

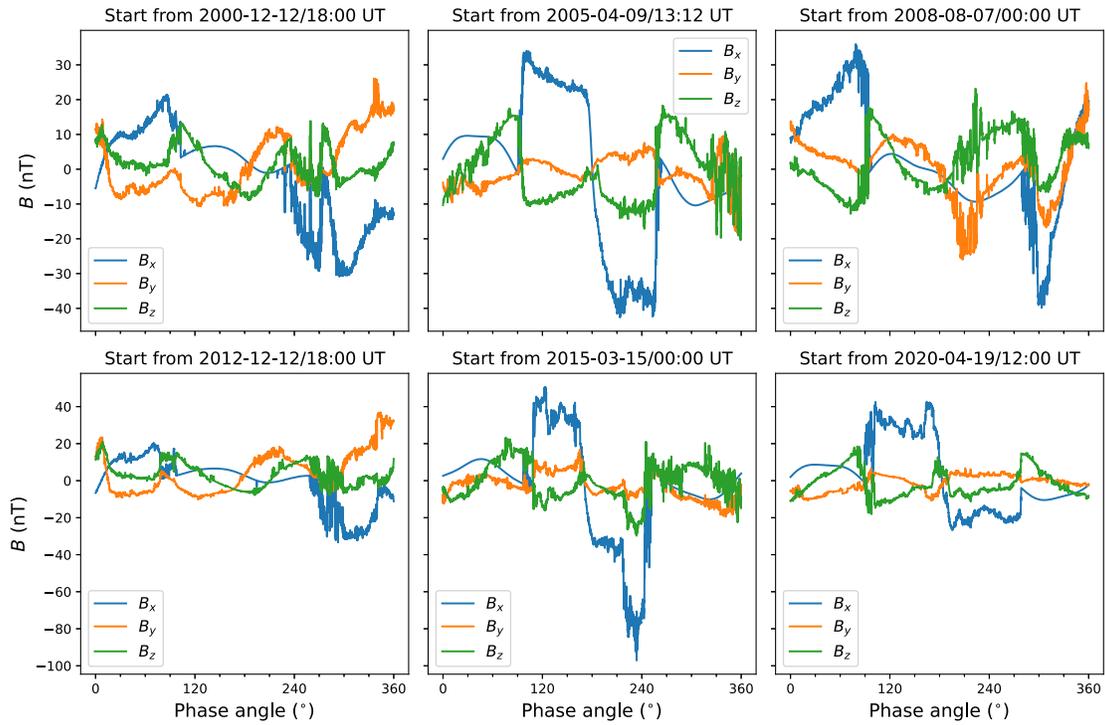

FIG. 3. Magnetic field along the TQ orbit obtained from the Tsyganenko model. The six panels correspond to the magnetic field distributions along each orbit of the TQ satellite in the six cases.

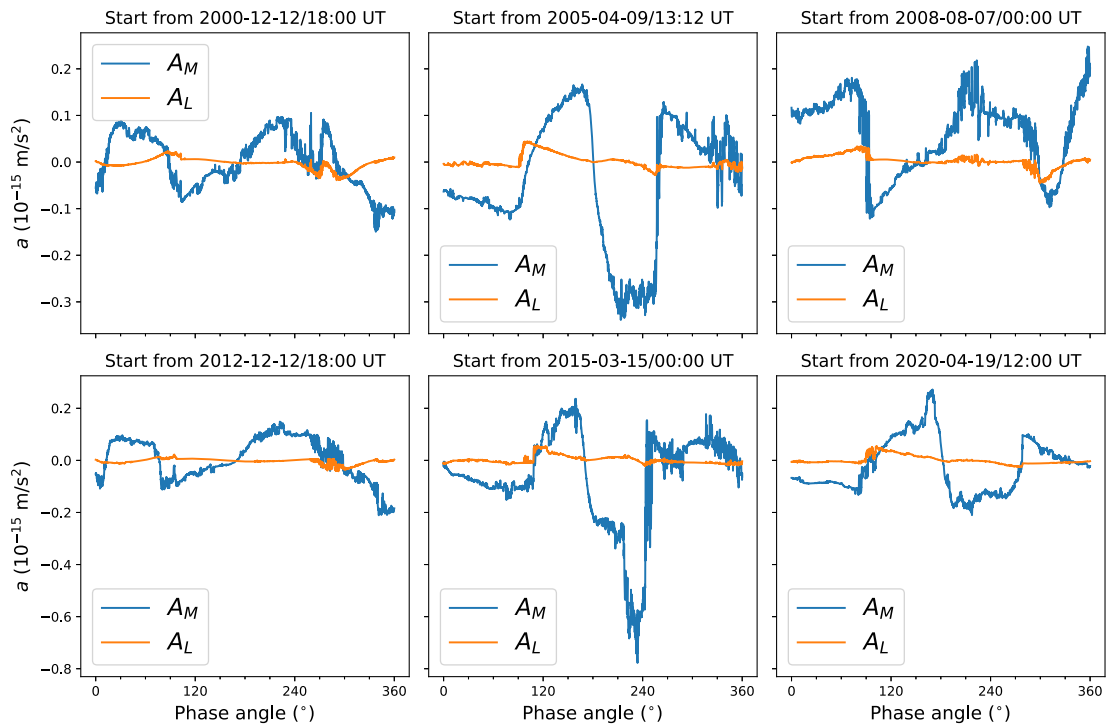

FIG. 4. Distributions of the acceleration noises along TQ orbit in the six cases. Blue curves correspond to $A_\mathrm{M}$, and orange curves correspond to $A_\mathrm{L}$.

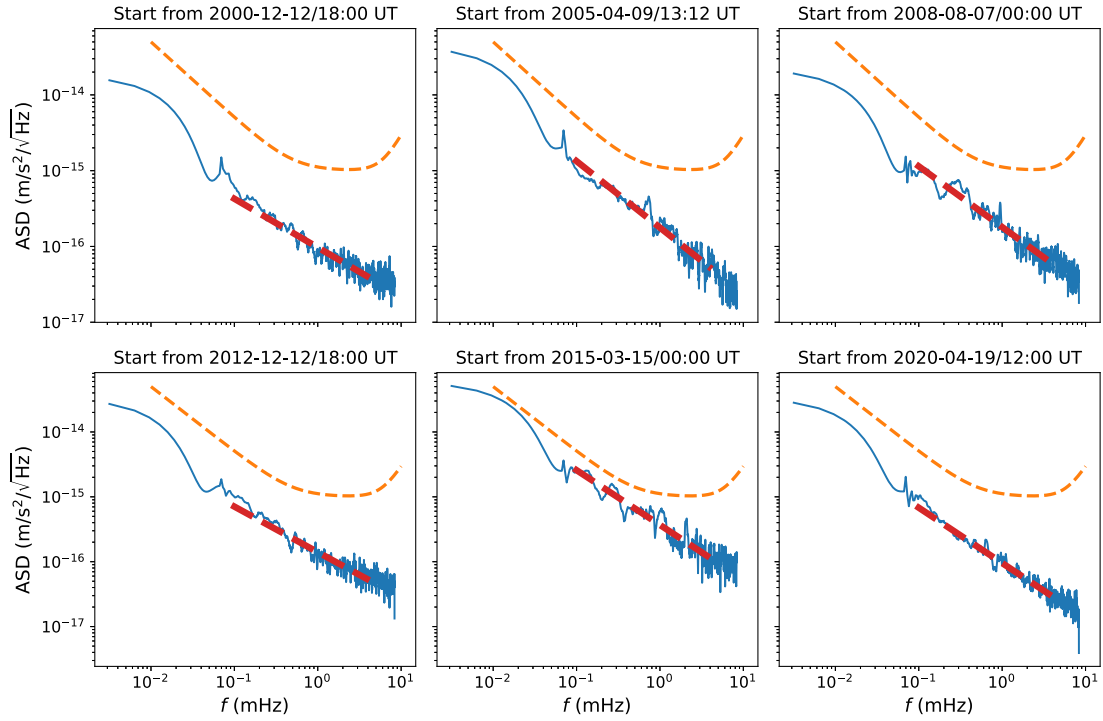

FIG. 5. ASDs of $A_M$ in the six cases. Blue curves are the ASDs of $A_M$. Red dashed lines are the best fits of the ASDs of $A_M$. Orange dashed curves are the requirement of the acceleration noise for TQ.

The ASD of TQ's acceleration requirement $\sqrt{S_a}$ is expressed as follows [71]:

$$\sqrt{S_a} = 1 \times \sqrt{1 + \left(\frac{f_{c1}}{f}\right)^2}\sqrt{1+\left(\frac{f}{f_{c2}}\right)^2}\ \mathrm{fm}/(\mathrm{s}^2\,\mathrm{Hz}^{1/2}),\tag{6}$$

where the transfer frequencies $f_{c1} = 0.5$ mHz and $f_{c2} = 6$ mHz. The requirement of acceleration noise for TQ is shown as the orange curve in Fig. 5. Here we assign the ratio between $A_{1\,\mathrm{mHz}}$ and $\sqrt{S_a}$ as $R_{1\,\mathrm{mHz}} = A_{1\,\mathrm{mHz}}/\sqrt{S_a}$. $R_{1\,\mathrm{mHz}}$ for the six cases are 0.086, 0.158, 0.161, 0.125, 0.329, and 0.085, respectively; The power indexes $\alpha$ of the fitting results for the six cases are $-0.642$, $-0.877$, $-0.801$, $-0.701$, $-0.843$, and $-0.852$, respectively.

We obtain the fitted results of the ASDs of the 2300 cycles. As shown in Fig. 6, the blue dotted line represents the median for each frequency bin, and the 1-$\sigma$, 2-$\sigma$, and 3-$\sigma$ intervals of the 2300 fitted results are shaded orange, purple, and brown, respectively. Similar to $R_{1\,\mathrm{mHz}}$, we also assign the ratio between $A_M$ and $\sqrt{S_a}$ at 6 mHz as $R_{6\,\mathrm{mHz}}$. The mean and standard deviation values of $R_{1\,\mathrm{mHz}}$ and $R_{6\,\mathrm{mHz}}$ are $0.123 \pm 0.052$ and $0.028 \pm 0.013$, respectively. The cumulative distribution function (CDF) of the acceleration noise caused by space magnetic field is shown in Fig. 7. The left and right panels are the CDFs of $R_{1\,\mathrm{mHz}}$ and $R_{6\,\mathrm{mHz}}$, respectively. The blue and orange bins represent the CDF and reversed CDF. From the reversed CDF bins (orange bins), we get the occurrence probabilities of $R_{1\,\mathrm{mHz}} > 0.2$ and $> 0.3$ are only 7.9% and 1.2%, respectively, and the occurrence probability of $R_{6\,\mathrm{mHz}} > 0.2$ is zero. It indicates that the acceleration noise due to the space magnetic field exceeding 20% and 30% of TQ's requirement is rare. The budget of acceleration noise due to the space magnetic field is roughly set at 30% of the total acceleration noise at the current stage. The mean values of ASDs of acceleration noise due to space magnetic field are about 40% and 10% of the magnetic acceleration noise budget at 1 and 6 mHz, respectively; and in the vast majority of cases, the space magnetic acceleration noises do not exceed the current magnetic acceleration noise budget.

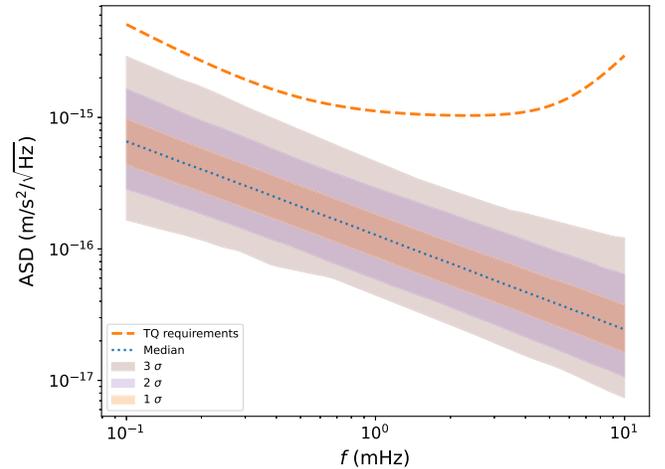

FIG. 6. Fitted results of ASDs of $A_M$ for 2300 cycles of a TQ satellite around the Earth in the frequency range of 0.1–10 mHz. The orange dotted curves are the requirement of acceleration noise for TQ. The blue line represents the median for each frequency bin. The 1-$\sigma$, 2-$\sigma$, and 3-$\sigma$ intervals of the 2300 fitted results are shaded orange, purple, and brown, respectively.

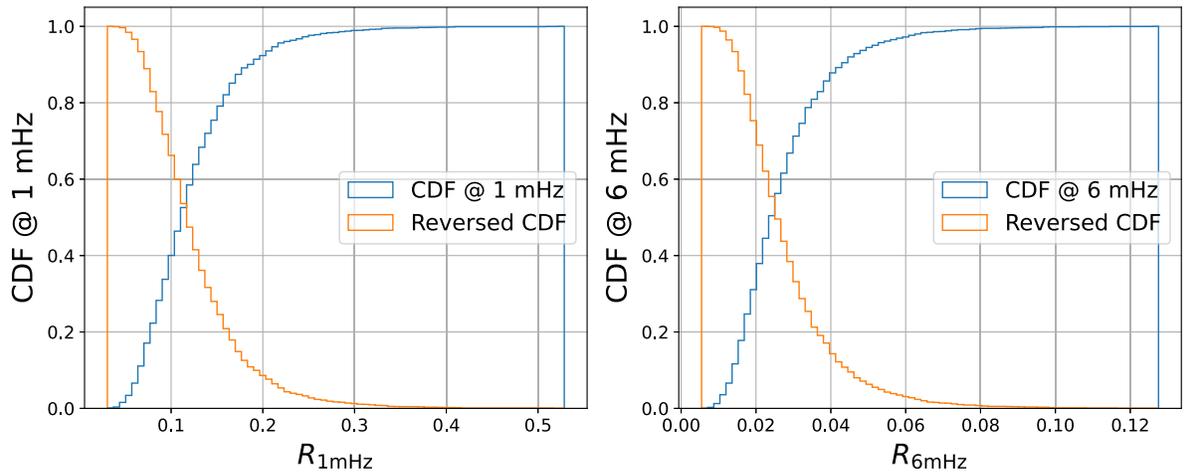

FIG. 7. CDF of the acceleration noise caused by the space magnetic field. The left and right panels are the CDFs of $R_{1\,\mathrm{mHz}}$ and $R_{6\,\mathrm{mHz}}$, respectively. The blue and orange bins are the cumulative occurrence probabilities and reversed cumulative occurrence probabilities, respectively.

### C. Comparison between the results of the SWMF and Tsyganenko models

Here, we compare the acceleration noises due to space magnetic field based on the SWMF and Tsyganenko models. Both models can characterize the magnetic structures of the magnetosphere, e.g., the magnetopause, magnetotail, current sheet, and cusp region. The parameter to be compared is the absolute value of $\boldsymbol{a}_{\mathrm{M1}}$ without the dc term [Eq. (4)], which is denoted as $|\boldsymbol{a}_{\mathrm{M1}}|$ here. The distributions of $|\boldsymbol{a}_{\mathrm{M1}}|$ in the four orbital scenarios ($\varphi_s = 0°$, $30°$, $60°$, and $90°$) based on the SWMF model are shown as the blue lines in the upper panels of Fig. 8, where $\varphi_s$ is the acute angle between the projection of the normal direction of TQ detector's plane on the ecliptic plane and the Sun-Earth direction. Then, we set the same inputs (magnetic field and $P_{\rm dyn}$) as in [30] for the Tsyganenko model, and calculate the magnetic field along the orbit for $\varphi_s = 0°, 30°, 60°$, and $90°$. Then, we calculate $|\boldsymbol{a}_{\rm M1}|$ for $\varphi_s = 0°, 30°, 60°$ and $90°$ based on the magnetic field obtained from the Tsyganenko model, and the results are shown as the orange dotted lines in the upper panels of Fig. 8. We can see that for $\varphi_s = 30°$, $60°$, and $90°$, $|\boldsymbol{a}_{\rm M1}|$ profiles obtained by the two space magnetic field models are in good agreement. Only for $\varphi_s = 0°$, there is a difference of $|\boldsymbol{a}_{\rm M1}|$ between the two models. Combined with Fig. 5 of Su *et al.* [30], for $\varphi_s = 30°, 60°$, and $90°$, when the TQ satellites are in the magnetotail (polar angle in the range from about $120°$ to $240°$), $|\boldsymbol{a}_{\rm M1}|$ obtained by two models match well with each other. When the TQ satellites are in the magnetosheath, the fluctuations of $|\boldsymbol{a}_{\rm M1}|$ obtained by the SWMF model are larger than those obtained by the Tsyganenko model. The magnetosheath of the Earth is the downstream of the bow shock, in which the disturbances of the parameters (e.g., the velocity, density, and magnetic field) are usually large [72]. In this region the SWMF model seems to perform better in characterizing the magnetosheath than the Tsyganenko model.

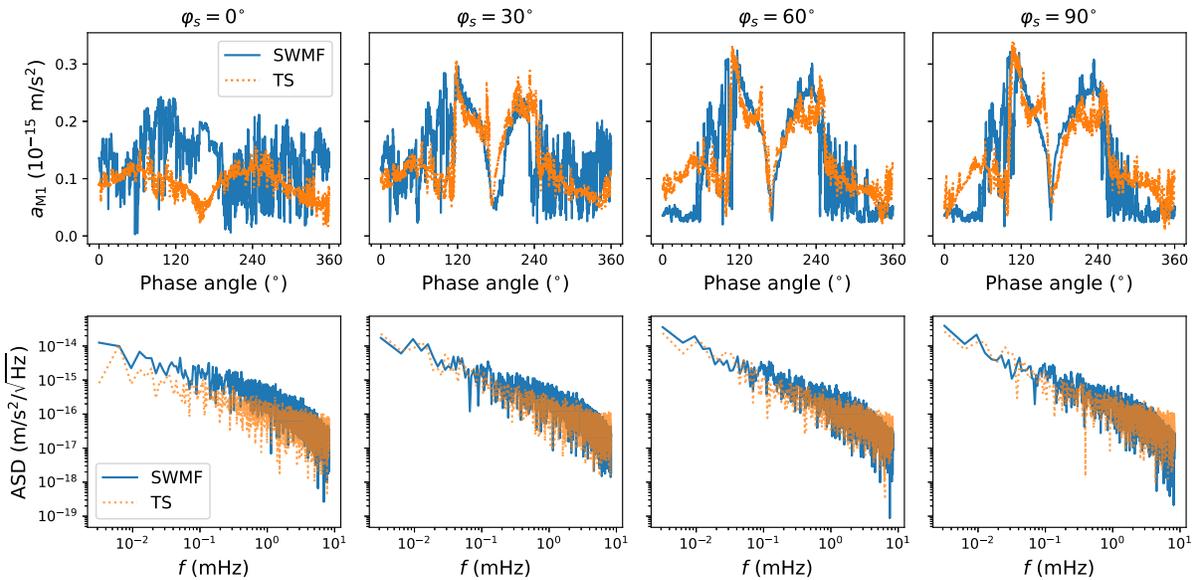

FIG. 8. The upper panels are the distributions of $|\boldsymbol{a}_{\mathrm{M1}}|$ along TQ orbit for $\varphi_s = 0°, 30°, 60°,$ and $90°$. The lower panels are the ASDs of $|\boldsymbol{a}_{\mathrm{M1}}|$ for $\varphi_s = 0°, 30°, 60°,$ and $90°$. The solid blue and dotted orange curves correspond to the SWMF and Tsyganenko models, respectively.

We further calculate the ASDs of $|\boldsymbol{a}_{\rm M1}|$ in the two space magnetic models. The ASDs of $|\boldsymbol{a}_{\rm M1}|$ for $\varphi_s = 0°, 30°, 60°$, and $90°$ based on the SWMF model are shown as the blue curves in the lower panels of Fig. 8, which are the same as the results in Fig. 8 of [30]. The ASDs of $|\boldsymbol{a}_{\rm M1}|$ for $\varphi_s = 0°, 30°, 60°$, and $90°$ based on the Tsyganenko model are shown as the orange dotted curves in the lower panels of Fig. 8. For the ASDs of the acceleration noises obtained by the Tsyganenko model, the spectral indices (slopes in log-log figure) are consistent over the entire frequency range. However, for the ASDs of the acceleration noises obtained by the SWMF model, the power-law spectrum can be divided into two parts separated at about 2 mHz, and the slope in the high frequency range $f \gtrsim 2$ mHz is steeper than that in low frequency range $f \lesssim 2$ mHz. The steepening of the slope in the high frequency range of the SWMF model makes its ASDs lower than that of the Tsyganenko model for all orbital cases. Such a drop is most probably due to the low spatial resolution of the SWMF magnetic model, which is 0.25 $R_{\rm E}$, i.e., $\sim 1600$ km. Considering the speed of the TQ satellites is about 2 km s$^{-1}$, the travel time across a grid mesh is 800 s, which implies that the spectrum above 1.2 mHz would be underestimated. With the temporal resolution of 60 s in our model, the spatial resolution of the SWMF ($\approx 1600$ km) is about one order of magnitude larger than what we need ($\approx 120$ km). So it requires interpolation when calculating $|\boldsymbol{a}_{\rm M1}|$, and the ASD of the interpolated data is likely to be lower than the actual one. In the frequency range $f \lesssim 2$ mHz, the ASDs of $|\boldsymbol{a}_{\rm M1}|$ in the SWMF have the spectra similar to those in the Tsyganenko model, although the amplitude of the Tsyganenko model is slightly smaller than that in the SWMF model when $\phi_s$ is closer to $0°$ as discussed in the previous paragraph. As the TQ satellites stay in the magnetosheath for a longer time, i.e., smaller $\varphi_s$, the Tsyganenko model underestimates the perturbations in the downstream of the bow shock, leading to lower ASDs obtained by the Tsyganenko model than by the SWMF model for small $\varphi_s$. The observations have verified that the spectral indices of the magnetic field in solar wind and magnetosphere are consistent in the frequency range from about $10^{-5}$ Hz to about 1 Hz [73–75]. It indicates that the spectral indices obtained by Tsyganenko model are in better agreement with observations than those by the SWMF model.

Both Tsyganenko and the SWMF models have been repeatedly validated in various scenarios [65,76,77]. However, the two models have their own limitations in the estimation of the acceleration noise due to space magnetic fields. The underestimation of the magnetic field perturbations in the bow shock downstream by the Tsyganenko model leads to the underestimation of the acceleration noises, while the limitation of the spatial resolution of the SWMF model leads to the underestimation of the acceleration noises in the high-frequency ($f \gtrsim 2$ mHz) range. The SWMF modol is an MHD model hence is too time-consuming, whereas the Tsyganenko model is a semiempirical model with much faster computational speed. Due to the limitation of the computational resources, the Tsyganenko model is suitable for the study of the statistical acceleration noises due to the space magnetic field on the timescale of tens of years with the temporal resolution of tens seconds.

## V. DISCUSSIONS

The orbit of TQ satellites will pass through the Earth's magnetosphere and the solar wind region, and the influence of the space plasma and magnetic field on the GWs detection should be considered [19,30]. The structure of the Earth's magnetosphere is affected by several parameters, such as the solar wind dynamic pressure, solar wind speed, plasma number density, interplanetary magnetic field, etc. Among all of the parameters, the primary one is the solar wind dynamic pressure, $P_{\rm dyn}$. While the magnetic pressure is dominant inside the Earth's magnetosphere, the dynamic pressure $P_{\rm dyn}$ is dominant in the solar wind. It was found very early that the Earth's magnetosphere is an approximate result of the balance between $P_{\rm dyn}$ in the solar wind and the Earth's magnetic pressure [78]. $P_{\rm dyn}$ can vary by several times in the timescale of minutes to hours. In contrast, the Earth magnetic dipole moment varies less than 0.1% per year [79]. That is to say, the Earth magnetic field is much more stable than $P_{\rm dyn}$. In this work, we investigated the influence of $P_{\rm dyn}$, which is the primary

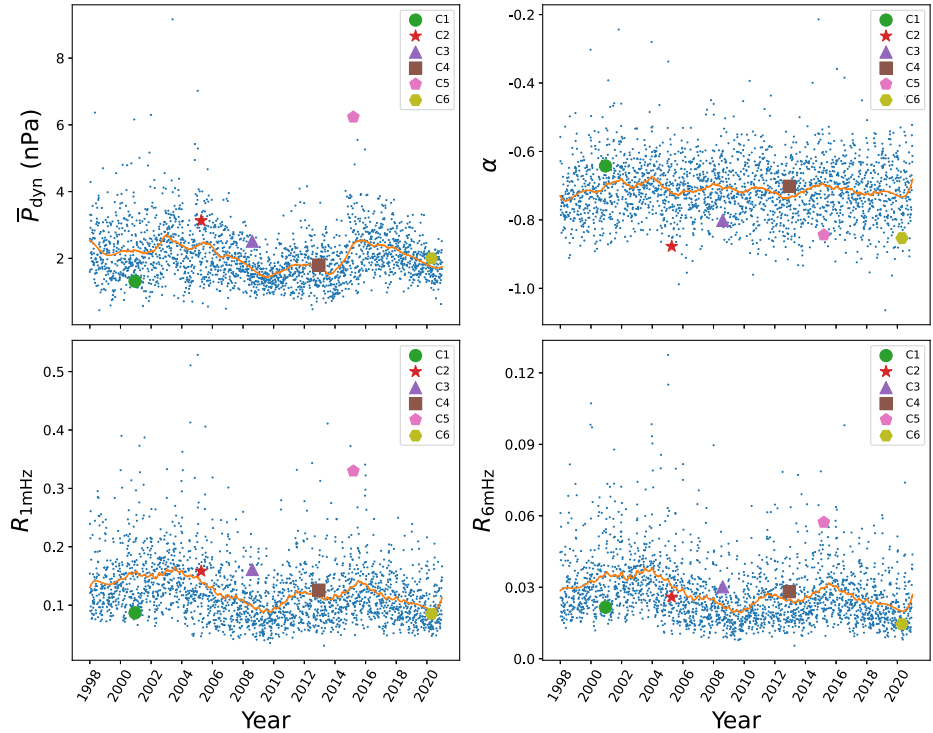

FIG. 9. The evolutions of $\bar{P}_{\text{dyn}}$, $R_{1\,\text{mHz}}$, $R_{6\,\text{mHz}}$, and $\alpha$ for the 2300 cycles of a TQ satellite around the Earth. C1–C6 mark the six cases.

parameter, on the acceleration noise due to space magnetic field for TQ.

Several solar activities, e.g., coronal mass ejections, corotating interaction regions, and interplanetary shocks, can significantly change $P_{\text{dyn}}$. In the timescale of tens of years, $P_{\text{dyn}}$ has a period of about 11 years, which is the same as the period of the solar cycle [80]. With $\bar{P}_{\text{dyn}}$ denoting the average value of $P_{\text{dyn}}$ in each orbital cycle (3.65 days) of a TQ satellite around the Earth, we calculated $\bar{P}_{\text{dyn}}$ for the 2300 orbital cycles of the satellite around the Earth, and the result is shown as scatterplots in the top left panel of Fig. 9, where the orange curve is the smoothed result of $\bar{P}_{\text{dyn}}$ by use of the Savitzky-Golay filter. There are two peaks around the years of 2004 and 2015, both of which are in the declining phase of the 11-year solar cycle. The $\bar{P}_{\text{dyn}}$ values of the six cases are represented by different symbols in the top left panel of Fig. 9, and their values are $1.314 \pm 0.395$ nPa, $3.131 \pm 1.765$ nPa, $2.515 \pm 2.243$ nPa, $1.796 \pm 0.441$ nPa, $6.233 \pm 4.379$ nPa, and $2.000 \pm 1.878$ nPa, respectively. The histogram of $\bar{P}_{\text{dyn}}$ is shown in Fig. 10, it indicates an asymmetric Gaussian distribution with a mean value of $2.07 \pm 0.78$ nPa.

The evolutions of $R_{1\,\text{mHz}}$ and $R_{6\,\text{mHz}}$ during the 2300 cycles are shown in the bottom panels of Fig. 9, the orange curves are the smoothed results of $R_{1\,\text{mHz}}$ and $R_{6\,\text{mHz}}$ by use of the Savitzky-Golay filter. The mean values of $R_{1\,\text{mHz}}$ and $R_{6\,\text{mHz}}$ are $0.123 \pm 0.052$ and $0.027 \pm 0.013$, respectively. Similar to the evolution of $\bar{P}_{\text{dyn}}$, there are two peaks around the years of 2004 and 2015 for both $R_{1\,\text{mHz}}$ and $R_{6\,\text{mHz}}$, implying that $\bar{P}_{\text{dyn}}$ can affect the intensity of the ASDs of acceleration noise caused by the space magnetic field. Moreover, we calculate the correlation coefficient between $\bar{P}_{\text{dyn}}$ and $R_{1\,\text{mHz}}(R_{6\,\text{mHz}})$ before and after smoothing, which are 0.483(0.390) and 0.757 (0.741), respectively. This again confirms the positive correlation between $\bar{P}_{\text{dyn}}$ and $R_{1\,\text{mHz}}(R_{6\,\text{mHz}})$. The positive correlation between $\bar{P}_{\text{dyn}}$ and $R_{1\,\text{mHz}}(R_{6\,\text{mHz}})$ can be understood as follows: In order to maintain the balance between the solar wind dynamic pressure and the magnetic pressure of the Earth's magnetosphere, the magnetic pressure of the Earth's magnetosphere will become larger with the increase of the solar wind dynamic pressure. Since the magnetic pressure is proportional to $B^2$, an increase of the magnetic pressure would lead to an increase of $|B|$. In turn, the increase of $|B|$ will lead to an increase of the amplitude of the acceleration noise due to the space magnetic field, which finally results in the increase of magnitude of the ASDs. The histogram of $R_{1\,\text{mHz}}$ is shown in Fig. 10, which, similar to the histogram of $\bar{P}_{\text{dyn}}$, reveals an asymmetric Gaussian distribution.

The evolution of $\alpha$ during the 2300 orbital cycles is shown in the top-right panel of Fig. 9, where the orange curve is the smoothed result of $\alpha$ by use of the Savitzky-Golay filter. The histogram of $\alpha$ is shown in Fig. 10, which indicates a symmetric Gaussian distribution with a mean value of $-0.712 \pm 0.086$. Unlike the evolutions of $\bar{P}_{\text{dyn}}$, $R_{1\,\text{mHz}}$, and $R_{6\,\text{mHz}}$, there is no significant periodicity for $\alpha$ over 23 years. Similarly, we calculate the correlation coefficient between $\bar{P}_{\text{dyn}}$ and $\alpha$ before and after smoothing, which are 0.114 and 0.235, respectively. This indicates that the correlation between $\bar{P}_{\text{dyn}}$ and $\alpha$ is not significant, and $\bar{P}_{\text{dyn}}$ does not affect the spectral index $\alpha$ of the ASDs of the acceleration noise due to the space magnetic field.

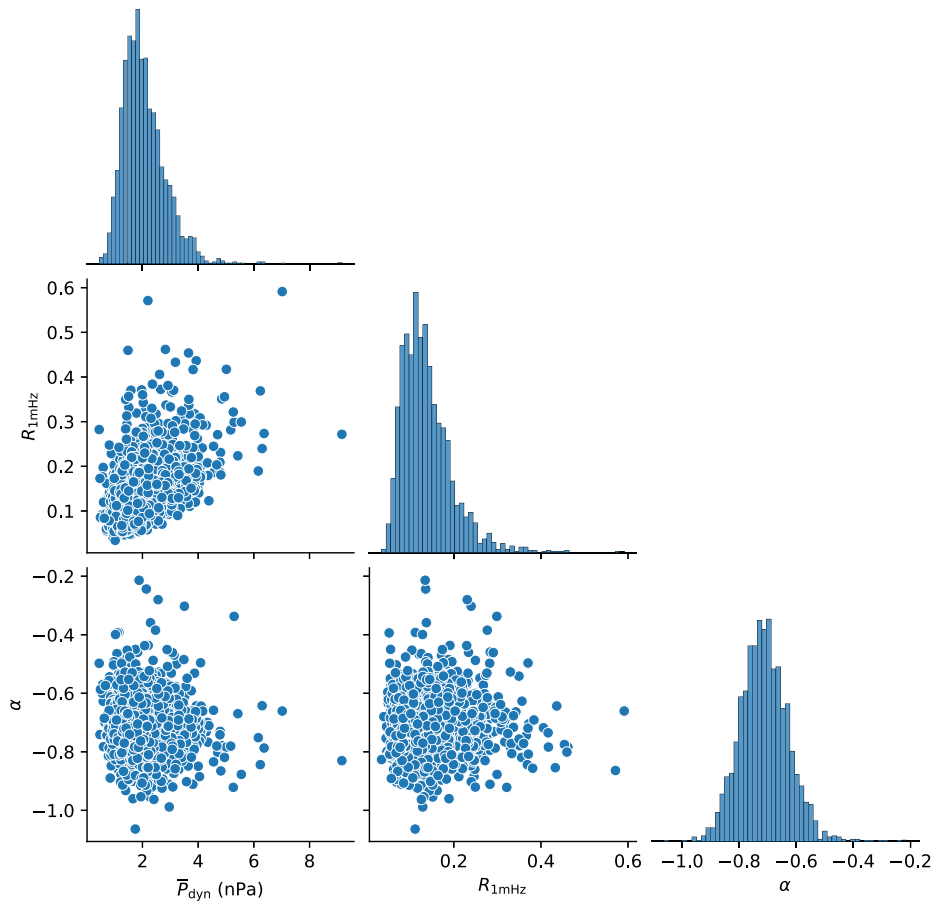

FIG. 10. Corner plots of parameters $\bar{P}_{\text{dyn}}$, $R_{1\,\text{mHz}}$, and $\alpha$ for 2300 cycles of a TQ satellite around the Earth.

It is noted that the time interval of the input conditions in this work covers 23 years (about two solar cycles), which consists of various types of interplanetary disturbances from high/low speed solar winds to coronal mass ejections. Although we do not separate each class of the heliospheric physical processes that have different impacts on the Earth (e.g., coronal mass ejections, corotating interaction regions, interplanetary shocks), the information of all these physical processes is contained in the input parameters of the Tsyganenko model (i.e., solar wind dynamic pressure, $B_z$ and $B_y$, and $D$st). Thus, the Earth's magnetic field during the 23 years obtained by the Tsyganenko model in this work is inclusive of the effects of all these heliospheric physical processes. The results of the acceleration noise due to the space magnetic field over the 23 years can provide useful information for TQ.

## VI. SUMMARY

In this work, we used a data-based empirical magnetic field model, the Tsyganenko model, to obtain the magnetic field distribution around the TQ's orbit over 23 years, which covers two solar cycles. Since the residual acceleration noise tolerability for TQ on the sensitive axis is two orders of magnitude lower than those on the nonsensitive axes, we focused on the acceleration along the sensitive axis in this work. With the obtained magnetic fields on the orbit, we calculated the distribution of the acceleration noise due to the space magnetic field along the sensitive axis for 2300 orbital cycles of a TQ satellite around the Earth. The results revealed that the residual acceleration has an asymmetric Gaussian distribution with mean values of $R_{1\,\text{mHz}}$ and $R_{6\,\text{mHz}}$ being $0.123 \pm 0.052$ and $0.027 \pm 0.013$. The correlation coefficient between the mean of solar wind dynamic pressure $\bar{P}_{\text{dyn}}$ and $R_{1\,\text{mHz}}(R_{6\,\text{mHz}})$ is 0.757(0.741), while the correlation coefficient between $\bar{P}_{\text{dyn}}$ and the spectral index ($\alpha$) of the ASD of residual acceleration is 0.235. These results indicate that $\bar{P}_{\text{dyn}}$ can affect the magnitude of the ASDs of the acceleration noise due to the space magnetic field, and the intensity of the acceleration noise increases with $\bar{P}_{\text{dyn}}$, but $\bar{P}_{\text{dyn}}$ has no significant effect on $\alpha$ of the acceleration noise.

The probabilities of occurrence of $R_{1\,\mathrm{mHz}} > 0.2$ and $> 0.3$ are only 7.9% and 1.2%, respectively, and the occurrence probability of $R_{6\,\mathrm{mHz}} > 0.2$ is zero.

## ACKNOWLEDGMENTS

S. W. is supported by the National Key R & D Program of China (No. 2020YFC2201200 and No. 2020YFC2200500). Y. W. gratefully acknowledges support from the National Key R & D Program of China (No. 2022YFC2205201), NSFC (No. 11973024), Major Science and Technology Program of Xinjiang Uygur Autonomous Region (No. 2022A03013-4), and Guangdong Major Project of Basic and Applied Basic Research (Grant No. 2019B030302001). Z. Z. B. is supported by NSFC (11727814 and 11975105). C. P. F. is supported by NSFC (12127901 and 11961131002). Z. C. is supported by the NSFC (No. 42074187) the National Key R & D Program of China (No. 2018YFC1503506). H. W. is supported by National Key R & D Program of China (2022YFC2204100 and 2021YFC2202500). Y. Y. is supported by NSFC (42204155) and Natural Science Foundation of Jiangsu, China (BK20210168).